\documentclass[floatfix,a4paper,prb,twocolumn,showkeys]{revtex4}

\usepackage{pstricks}
\usepackage{graphicx}
\usepackage{amsmath,amssymb}

\begin{document} 

\title{The protein folding network}

\author{Francesco Rao}
\author{Amedeo Caflisch}

\email[corresponding author, tel: +41 1 635 55 21,
fax: +41 1 635 68 62, e-mail: ]{caflisch@bioc.unizh.ch}
	     
\affiliation{Department of Biochemistry, University of Zurich, 
	     Winterthurerstrasse 190, CH-8057 Zurich, Switzerlandi\\
             tel: +41 1 635 55 21, fax: +41 1 635 68 62, 
	     e-mail: caflisch@bioc.unizh.ch}


\begin{abstract}

The conformation space of a 20-residue antiparallel $\beta$-sheet peptide,
sampled by molecular dynamics simulations, is mapped to a network.
Conformations are nodes of the network, and the transitions between them are
links. The conformation space network describes the significant free energy
minima and their dynamic connectivity without projections into arbitrarily
chosen reaction coordinates.  As previously found for the Internet and the
World-Wide Web as well as for social and biological networks, the conformation
space network is scale-free and contains highly connected hubs like the native
state which is the most populated free energy basin.  Furthermore, the native
basin exhibits a hierarchical organization which is not found for a random
heteropolymer lacking a predominant free-energy minimum. The network topology
is used to identify conformations in the folding transition state ensemble, and
provides a basis for understanding the heterogeneity of the transition state
and denaturated state ensemble as well as the existence of multiple pathways.

\end{abstract}


\maketitle

Proteins are complex macromolecules with many degrees of freedom.  To fulfill their
function they have to fold to a unique three-dimensional structure (native
state).  Protein folding is a complex process governed by noncovalent
interactions involving the entire molecule. Spontaneous folding in a time range
of microseconds to seconds \cite{Daggett:Is} can be reconciled with the large
amount of conformers by using energy landscape analysis
\cite{Bryngelson:Random,Leopold:Protein,Karplus:The}.  The main difficulty of
this analysis is that the free-energy has to be projected on arbitrarily chosen
reaction coordinates (or order parameters).  In many cases a simplified
representation of the free-energy landscape is obtained where important
informations on the non-native conformation ensemble and the folding transition
state ensemble are hidden. Moreover, the possible transitions between
free-energy minima cannot be displayed in such projections which hinder the
study of pathways and folding intermediates. The characterization of the
free-energy minima and the connectivity among them, i.e., possible transitions
between minima, for peptides and proteins is still an unresolved problem.

In the last five years many complex systems, like the World-Wide Web, metabolic
pathways, and protein structures have been modeled as networks
\cite{Jeong:The,Albert:Diameter,greene:protein}. Intriguingly, common
topological properties have emerged from their organization \cite{newman:siam}.
A description of the potential energy landscape without the use of any
projection has been given in terms of networks for a Lennard-Jones cluster of
atoms \cite{Doye:scalefree}.  

Here, we introduce complex network analysis \cite{newman:siam} to study the
conformation space and folding of beta3s, a designed 20-residue sequence whose
solution conformation has been investigated by NMR spectroscopy
\cite{DeAlba:Denovo}. The NMR data indicate that beta3s in aqueous solution
forms a monomeric (up to more than 1mM concentration) triple-stranded
antiparallel $\beta$-sheet (Fig.\ 1, bottom), in equilibrium with the
denaturated state \cite{DeAlba:Denovo}. We have previously shown that in
implicit solvent \cite{Ferrara:Evaluation} molecular dynamics simulations
beta3s folds reversibly to the NMR solution conformation, irrespective of the
starting conformation \cite{Ferrara:Folding,Cavalli:Weak}.  
We consider conformations sampled by molecular dynamics simulations and the
transitions between them as the network nodes and links, respectively.  
The network analysis allows to identify the topological properties that are
common to both beta3s, which folds to a unique three-dimensional structure
\cite{Cavalli:Fast,DeAlba:Denovo}, and a random heteropolymer which lacks a
single preferential conformation like the native state despite it has the same
residue composition as beta3s. These properties include the presence
of several free-energy minima and highly connected conformations (hubs).  On the other
hand, a hierarchical modularity \cite{Bara:Hier} in the proximity of the native
state is peculiar of a folding sequence.

\begin{figure}[h]
\includegraphics[angle=0,width=80mm]  {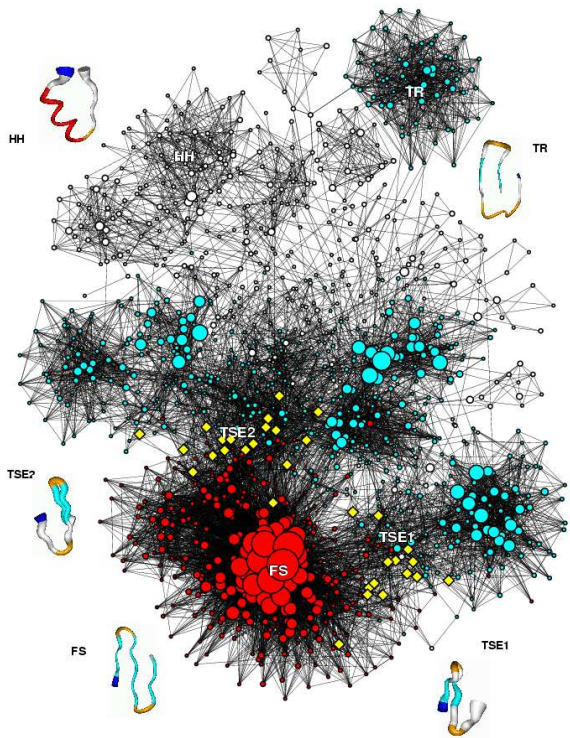}
\caption{Beta3s conformation space network.  The size and color
coding of the nodes reflect the statistical weight $w$ and average neighbor
connectivity $k_{nn}$, respectively.  White, cyan, and red nodes have
$k_{nn}<30$, $30 \leq k_{nn} \leq 70$, and $k_{nn}>70$, respectively.
Representative conformations are shown by a pipe colored according to secondary
structure: white stands for coil, red for $\alpha$-helix, orange for bend, cyan
for strand and the N-terminus is in blue. The variable radius of the pipe
reflects structural variability within snapshots in a conformation.  The yellow
diamonds are folding TS conformations (TSE1, TSE2, see text for details)
characterized by a connectivity/weight ratio $k/2\tilde w>0.3$, a clustering
coefficient $C<0.3$, and $60<k_{nn}<80$. This figure was made using
\textit{visone} (www.visone.de) and \textit{MOLMOL}\cite{Koradi96}
visualization tools.}

\end{figure}

\section{Model and Methods} 

\vspace{0.0cm}\noindent{\bf Molecular dynamics simulations}\hspace{0.2cm} The
simulations and part of the analysis of the trajectories were performed with
the program CHARMM {\cite{Brooks:CHARMM}}.  Beta3s was modeled by explicitly
considering all heavy atoms and the hydrogen atoms bound to nitrogen or oxygen
atoms (PARAM19 force field {\cite{Brooks:CHARMM}}).  A mean field approximation
based on the solvent accessible surface was used to describe the main effects
of the aqueous solvent on the solute \cite{Ferrara:Evaluation}.  The two
parameters of the solvation model were optimized without using beta3s.  The
same force field and implicit solvent model have been used recently in
molecular dynamics simulations of the early steps of ordered aggregation
\cite{Gsponer:Therole}, and folding of structured peptides ($\alpha$-helices
and $\beta$-sheets) ranging in size from 15 to 31 residues
\cite{Ferrara:Evaluation,Ferrara:Folding,Hiltpold:Free}, as well as small
proteins of about 60 residues \cite{Gsponer:Role,Gsponer:Molecular}. Despite
the absence of collisions with water molecules, in the simulations with
implicit solvent the separation of time scales is comparable with that observed
experimentally.  Helices fold in about 1 ns \cite{Ferrara:Thermodynamics},
$\beta$-hairpins in about 10 ns \cite{Ferrara:Thermodynamics} and
triple-stranded $\beta$-sheets in about 100 ns \cite{Cavalli:Weak}, while the
experimental values are $\sim$0.1 $\mu$s \cite{Eaton:Fast}, $\sim$1 $\mu$s
\cite{Eaton:Fast} and $\sim$10 $\mu$s \cite{DeAlba:Denovo}, respectively.
Recently, four molecular dynamics simulations of beta3s were performed at 330 K
for a total simulation time of 12.6 $\mu$s \cite{Cavalli:Fast}.  There are 72
folding events and 73 unfolding events and the average time required to go from
the denatured state to the folded conformation is 83 ns.  The 12.6 $\mu$s of
simulation length is about two orders of magnitude longer than the average
folding or unfolding time, which are similar because at 330 K the native and
denatured states are almost equally populated \cite{Cavalli:Fast}. For the
network analysis the first 0.65 $\mu$s of each of the four simulations were
neglected so that along the 10 $\mu$s of simulations there are a total of $5
\times 10^{5}$ snapshots because coordinates were saved every 20 ps.  The
sequence of the random heteropolymer is a randomly scrambled version of the
beta3s sequence with the same residue composition.  It was simulated for 2
$\mu$s and $10^{5}$ snapshots were saved.  The conditions for the molecular
dynamics simulations, i.e., force field, solvation model, temperature, and time
interval between saved snapshots were the same for both peptides.   

\vspace{0.5cm}\noindent{\bf Construction of the protein folding
network}\hspace{0.2cm} To define the nodes and links of the network the
secondary structure was calculated \cite{DSSP:cont} for each snapshot
(Cartesian coordinates of the atomic nuclei) saved along the molecular dynamics
trajectory.  A "conformation" is a single string of secondary structure
\cite{DSSP:cont}, e.g., the most populated conformation for beta3s (FS in Fig.\
1) is {\tt -EEEESSEEEEEESSEEEE-} where "{\tt E}", "{\tt S}", and "{\tt -}"
stand for extended, turn, and unstructured, respectively.  There are 8 possible
"letters" in the secondary structure "alphabet". Since the N- and C-terminal
residues are always assigned an "{\tt -}" \cite{DSSP:cont} a 20-residue peptide
can assume $8^{18}\simeq 10^{16}$ conformations.  Conformations are nodes of
the network and the transitions between them are links. A weight $\tilde w$ is
assigned to each node to take into account the free-energy of each conformation
and is equal to the number of snapshots with a given secondary structure
string.  The statistical weight $w$ of a node is equal to the weight normalized
by the total number of snapshots in the simulations ($5 \times 10^{5}$ and
$10^{5}$ for beta3s and the random heteropolymer, respectively).  Considering
all the conformations visited during a $\mu s-scale$ simulation can yield to a
computationally intractable network size. For this reason we used for the
network analysis the 1287 conformations of beta3s with significant weight
($\tilde w\geq 20$ per conformation).  Two nodes are connected by an undirected
link (and called neighbors) if they either include a pair of snapshots that are
visited within 20 ps or they are separated by one or more conformations with
less than 20 snapshots each.  For the 2 $\mu$s of the random heteropolymer a
threshold of $\tilde w\geq 4$ was used, so that $w\geq 4\times 10^{-5}$ as in
the beta3s network. The choice of a threshold value is somewhat arbitrary but
the network properties are robust for a large range of threshold values (see
Supplementary material).  

The properties of the network are robust also with respect to the length of the
simulation time and the definition of the nodes.  The topological properties
are independent from simulation lengths if one considers more than $2\ \mu s$.
The correlation between statistical weight and connectivity, as well as
power-law behavior of the connectivity distribution and $1/k$ behavior of the
clustering coefficient distribution (see below) are essentially identical after
2, 4, and $10\ \mu s$.  As an example, the exponent of the power-law is $2.0$
for the beta3s networks based on $2$, $4$ and $10\ \mu s$ of simulation time.
Defining nodes by grouping snapshots according to root mean square deviations
({\sc rmsd}) in coordinates of C$_{\alpha}$-C$_{\beta}$ atoms yields the same
overall properties i.e., power-law distribution of the links (with similar
$\gamma$ value) and $1/k$ tail of the clustering distribution.  Grouping
snapshots according to secondary structure motifs does not require the use of
an arbitrarily chosen {\sc rmsd} cutoff, and is able to capture the
fluctuations of partially structured conformations\cite{DSSP:cont}.

\vspace{0.5cm}\noindent{\bf Evaluation of $\bf P_{fold}$}\hspace{0.2cm} The TS
ensemble can be defined as the set of structures which have the same
probability of folding ($P_{fold}$) or unfolding in trajectories started with
varying initial conditions\cite{Du:Tcoor}.  For each putative TS conformation,
the probability to fold before unfolding was calculated by 100 very short
trajectories at 330~K started from ten snapshots within a node.  The only
difference between the ten runs was the seed for the random number generator
used for the initial assignment of the atomic velocities.  A trajectory was
considered to lead to folding (unfolding) if it visits first structures with a
fraction of native contacts $Q>22/26$ ($Q<4/26$) \cite{Ferrara:Folding}.

\section{Results and Discussion}

To study the conformation space network of polypeptides we concentrate
on the analysis of topology, i.e., on the study of the connectivity between
different conformations, leaving for a later study the analysis of transition
rates. We have investigated the network topologies of several peptides but on
this paper we focus on beta3s and the random scrambled version of it.
Additional details can be find in the Supplementary material where the network
properties of another structured peptide and a glycine homopolymer are
presented.

\vspace{0.5cm}\noindent{\bf Conformation space network of a structured
peptide}\hspace{0.2cm} The conformation space network and relevant structures
of beta3s are shown in Fig.\ 1.  The group of nodes at the bottom of Fig.\ 1
(red nodes) represents the native state basin (FS).  The native basin is
connected to a wide region of nodes with significant native content (cyan
circles in the middle of Fig.\ 1).  Although many heterogeneous routes can be
taken to reach the folded state (in agreement with lattice
simulations\cite{Onuchic:TSE,Schonbrun:Fast}), most of the folding events have
common structural features that define two average folding pathways.  The less
frequented average pathway (see Ref \cite{Cavalli:Weak} but also the density of
transitions in Fig\ 1 bottom right) consists of conformations that have the
N-terminal hairpin formed while the C-terminal strand is mostly unstructured
with non-native hydrogen bonds at the turn (TSE1 in Fig.\ 1). The second and
most frequented average pathway includes conformations with a well formed
C-terminal hairpin while the N-terminal strand is disordered (TSE2 in Fig.\ 1),
namely it can be out-of-register or mostly unstructured.  It is interesting to
note that the same two folding pathways were observed experimentally for a
24-residue peptide with the same folded state as
beta3s\cite{Griffiths:Structure}.  Furthermore multiple folding pathways have
recently been detected by kinetic analysis of a $\beta$-sandwich
protein\cite{Wright:Parallel}.

The denatured state ensemble is very heterogeneous and includes high enthalpy,
high entropy conformations (e.g., the partially helical conformations, denoted
HH in Fig.\ 1) but also low enthalpy, low entropy conformations (e.g., the
curl-like trap, TR).  The former are loosely linked clusters of conformations
with similar secondary structure (see Tab.\ 1) which are characterized by an
unfavorable effective energy (sum of peptide
potential energy and solvation energy) and fluctuating unstructured residues
(e.g., the terminal of the helix shown on top left of Fig.\ 1).  On the
contrary, low enthalpy, low entropy traps form tightly linked clusters with
almost identical secondary and tertiary structure, favorable effective
energy (similar to the one of the native structure, see Tab.\ 1) and no
fluctuating residues (e.g., Fig.\ 1, top right).  Taken together, these results
indicate that FS is entropically favored over low enthalpy conformations like
TR, i.e., FS has more flexibility than TR.  A possible explanation is that the
C-terminal carboxy is involved in four hydrogen bonds in TR (with the backbone
NH's of residues 4-7), whereas both termini undergo rather large fluctuations
in FS.  In addition, a more favorable van der Waals energy in TR is consistent
with a denser packing in TR than in FS.  

\setcounter{table}{0}
\renewcommand{\thetable}{\arabic{table}}

\begingroup
\squeezetable
\begin{table}[h]
\begin{ruledtabular}
\caption{Energetic comparison of folded and denaturated state.  The free-energy
of conformation $i$ is \mbox{${\cal F}_i = -k_BT \log(w_i)$}, where $w_i$ is the
probability along the trajectory to find the peptide in the conformation $i$.}
\begin{tabular}{ccc}
\parbox{4.0cm}{\flushleft{\bf Folded state (FS)}}   
& $\left < {\cal E} \right >$ \footnote{Average effective energy} 
& $\Delta{\cal F}$ \footnote{Free-energy relative to the most populated
conformation. All values are in kcal/mol.  The conformational entropy of the
peptide is equal to $(\left < {\cal E} \right > - {\cal F})/T$. Note that the
curl-like traps are entropically penalized with respect to the native state.}\\
{\tt      -EEEESSEEEEEESSEEEE-} &     -7.6 &      0\\
{\tt      -EEE-STTEEEEESSEEEE-} &     -8.6 &      0.1\\
{\tt      -EEEESSEEEEE-STTEEE-} &     -8.4 &      0.5\\
{\tt      -EEE-STTEEEE-STTEEE-} &     -9.2 &      0.7\\
\parbox{4.0cm}{\flushleft{\bf Helical conformations (HH)}} &   &         \\
{\tt      ---HHHHHHHHHHS------} &      0.9 &      3.1\\
{\tt      -HHHHHHHHHHHHS------} &     -1.9 &      3.3\\
{\tt      ---HHHHHHHHHHTT-----} &      0.7 &      3.5\\
{\tt      ---HHHHHHHHHH-------} &      0.5 &      3.7\\
{\tt      -HHHHHHHHHHHHTT-----} &     -0.8 &      3.7\\
{\tt      --TT--HHHHHHHHHHHHH-} &     -0.8 &      3.8\\
\parbox{4.0cm}{\flushleft{\bf Curl-like trap (TR)}} &          &         \\
{\tt      ---SSGGG-EEE-STTTEE-} &     -7.8 &      3.4\\
{\tt      ---SSSS--EEE-STTTEE-} &     -7.0 &      3.5\\
{\tt      ---S-GGG-EEE-STTTEE-} &     -9.3 &      3.7\\
{\tt      ---SSGGG-EEE-SGGGEE-} &     -9.6 &      3.7\\
{\tt      ---SSTTT-EEE-STTTEE-} &     -8.4 &      3.7\\
\end{tabular}
\end{ruledtabular}
\end{table}
\endgroup

Note that the network description of non-native conformations is more detailed
than the one obtained by projecting the free energy surface on progress
variables (e.g., based on fraction of native contacts).  In such projections,
for low values of the fraction of native contacts structures as diverse as
helices and the curl-like conformations mentioned above are not
distinguished.  Even the ensemble with half of the native contacts is
heterogeneous and hard to classify. Using as reaction coordinate the {\sc rmsd}
(with respect to a given structure) or the radius of gyration is even less
selective.  Only when a clever combination of variables is used it is possible
to have a more detailed description of the free-energy landscape.  The network
description of the conformation space gives a synthetic and systematic view of
all the possible conformations accessed by the system and their transitions.
By considering the statistical weight of the nodes a thermodynamical
description of the system is obtained.

\begin{figure}[h!]
\includegraphics[angle=-90,width=80mm]{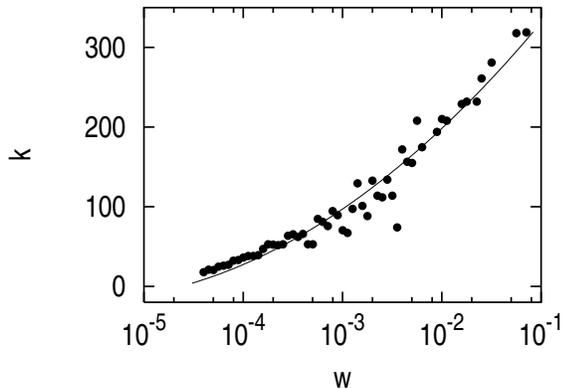}
\caption{Correlation between the statistical weight $w$ and the
connectivity $k$ for beta3s.  The connectivity is proportional to $log^2(w)$
with a correlation coefficient of 0.88 (solid line).  The correlation and the
fit are calculated over all nodes of the network but in the figure logarithmic
binning is applied to reduce noise.}
\end{figure}

The high correlation between the statistical weight of a node and its number of
links (Fig.\ 2) shows that the most connected nodes are also low lying minima
on the free-energy landscape.  This indicates that the conformation space
network describes the significant free energy minima and their dynamic
connectivity, without projection, where highly populated nodes are minima of
free-energy and the set of nodes densely connected to them make up the basins
of such minima.

\vspace{0.5cm}\noindent{\bf Folding and network topology}\hspace{0.2cm} The
average neighbor connectivity $k_{nn}$ of beta3s (Fig.\ 3a), i.e., the average
number of links of the neighbors of a given node, is rather heterogeneous,
highlighting the presence of different connection rules in different regions of
the network.  This is not the case for the random heteropolymer (Fig.\ 3b)
whose basins have organization and statistical weight similar among each others
as previously found for most homopolymers \cite{Vendrusc:SW}.  Note that for
beta3s the native state is well discriminated by $k_{nn}$ (red nodes in Fig.\ 1
and top band in Fig.\ 3a).

The connectivity distribution of conformation space networks shows a well
pronounced power-law tail $P(k)=k^{-\gamma}$ with $\gamma=2.0$ for both beta3s
and the random heteropolymer (Fig.\ 4a) as well as another structured peptide
\cite{demarest:alphalac} and homoglycine, i.e.\ (Gly)$_{20}$ (see Supplementary
material). The power-law is due to the presence of few largely connected "hubs"
while the majority of the nodes have a relatively small number of
links\cite{Bara:Scale}. This behavior has been previously observed for several
biological\cite{Jeong:The}, social\cite{Newman:Scicol} and technological
networks\cite{Albert:Diameter}, which in the literature take the name of
scale-free networks. In terms of free-energy this means that only a few low
lying minima are present but they act as "hubs" with a large number of routes
to access them.

The average clustering coefficient $C$ is a measure of the probability that any
two neighbors of a node are connected.  Beta3s and the heteropolymer have $C$
values of $0.49$ and $0.28$, respectively. These values are one order of
magnitude larger than random realizations of the two networks with the same
amount of nodes and links. The native basin of beta3s includes the nodes with
the largest number of links of the network.  These nodes give rise to the $1/k$
tail of the clustering distribution (Fig.\ 4b), i.e., an inherently
hierarchical organization\cite{Bara:Hier} of the conformations in the native
basin of beta3s.  Such organization is not observed for the non-native region
of beta3s and the random heteropolymer. Note that the power-law scaling of the
connectivity distribution can be considered as a general property of
free-energy landscapes of polypeptides, whereas a hierarchical organization of
the nodes reflects a single pronounced free-energy basin of attraction (like
the native state).  

\begin{figure}[h!]
\includegraphics[angle=-90,width=90mm]{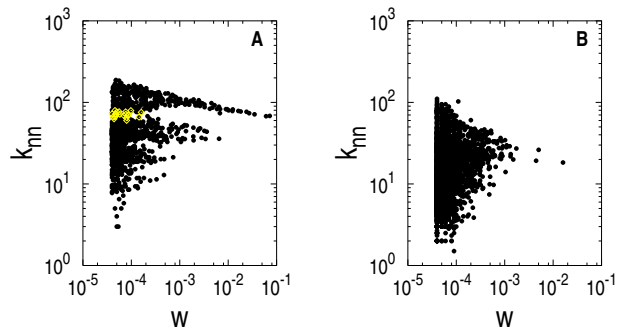}
\caption{Average neighbor connectivity $k_{nn}$ plotted as a function of the
statistical weight for the 1287 nodes of beta3s (A) and for the 2658 nodes of
the random heteropolymer (B). $k_{nn}$ of node $i$ is the average number of
links of the neighbors of node $i$. The yellow diamonds are folding transition
state conformations (see also Fig.\ 1 and text) characterized by a
connectivity/weight ratio $k/2\tilde w>0.3$, a clustering coefficient $C<0.3$,
and $60<k_{nn}<80$.}
\end{figure}

\begin{figure}[h] \includegraphics[angle=-90,width=90mm]{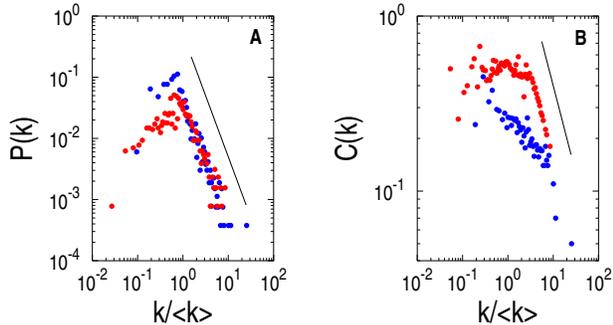}
\caption{Topological properties of conformation space networks. Red and
blue data points are plotted for beta3s and a random heteropolymer,
respectively. For a direct comparison, the connectivity $k$ is normalized by
the average connectivity $\left<k\right>$ of each network. Logarithmic binning
is applied to reduce noise.  (A) The connectivity distribution $P(k)$ is the
probability that a node (conformation) has $k$ links (neighbor conformations).
The straight line corresponds to a power-law fit $y=x^{-\gamma}$ on the tail of
the distribution with $\gamma=2.0$. (B) The clustering coefficient $C$
describes the cliques of a node. For node $i$ it is defined as
$C_i=\frac{2n_i}{k_i(k_i-1)}$, where $k_i$ is the number of neighbors of node
$i$ and $n_i$ is the total number of connections between them. Values of $C$
are averaged over the nodes with $k$ links. The straight line corresponds to a
power-law fit $y=x^{-1}$ on the tail of the distribution of beta3s.}
\end{figure}

\vspace{0.5cm}\noindent{\bf Transition state ensemble}\hspace{0.2cm} As
mentioned above folding is a complex process with many degrees of freedom
involved and it is difficult (or even not possible) to define a single
reaction coordinate to monitor folding events
\cite{Chan:Protein,Karplus:Aspects}.  Hence, it is very difficult to isolate
transition state (TS) conformations from equilibrium sampling.  The TS
conformations are saddle points, i.e., local maxima with respect to the
reaction coordinate for folding and local minima with respect to all other
coordinates.  For this reason we identified the nodes with a high
connectivity/weight ratio $k_i/2\tilde w_i$ and low clustering coefficient
value $C_i$ as putative TS conformations.  The former criterion guarantees that
these nodes are accessed and exited, most of the time, by a different route,
i.e., they can be directly reached from different conformations of the network
space.  The low clustering coefficient value guarantees that 
the neighbors of these
conformations are likely to be disconnected.  These two conditions are
necessary but not sufficient because they do not distinguish folding TS
conformations from saddle points between unfolded conformations.  
Since the folding TS conformations are linked to both nodes in the native state
(having large number of links) and in the
denatured state (small/intermediate number of links), we speculated that
folding TS conformations should have values of the average neighbor
connectivity $k_{nn}$ within a certain range.  For nodes with high
connectivity/weight ratio and low clustering
coefficient, a remarkable correlation of $0.89$ was found
between the average neighbor connectivity $k_{nn}$  and $P_{fold}$ (Fig.\ 5),
which is the probability of a given conformation to fold before
unfolding\cite{Du:Tcoor}.  A $P_{fold}$ value close to 0.5 is expected for
conformations on top of the folding TS barrier\cite{Gsponer:Molecular} and the
correlation suggests that network properties can be used to predict folding TS
conformations.  These are shown in Fig.\ 1 and 3a with yellow diamonds. As
discussed above two main average folding pathways are observed.  The less
frequented one is characterized by a transition state ensemble of conformations
with the first hairpin in a native form (residues 1-13) and a bend in
correspondence of the the second native turn (residues 14-15). The
C-terminal
residues form a straight structure with almost no contacts, either native or
non-native. The second average pathway shows a transition state with the second
native harpin formed (residues 7-20) and a bend in correspondence of the the
first native turn (residues 5-6). Such a symmetrical behavior is presumably due to the
simplicity and symmetry of the native conformation as well as the symmetry  
in the sequence (sequence identity  of 67\% between the two hairpins).
The folding TS conformations of beta3s form an heterogeneous ensemble with
C$_{\alpha}$ root mean square deviations within contributing structures between
3 and 6 \AA.  In contrast to previous molecular dynamics studies in which
progress variables based on fraction of native contacts were used to describe
TS conformations \cite{Lazaridis:New,Ferrara:Folding}, the network properties
yield a description of the folding TS ensemble (Fig.\ 1) which does not depend
on the choice of reaction coordinates.  Interestingly, the folding TS
conformations of beta3s have about one-half of the native contacts formed but
this is not a sufficient criterion (Table S1 in Supplementary material).
Moreover, there is no correlation between the fraction of native contacts and
the probability of folding.  As a control, $P_{fold}$ values smaller than 0.15
were obtained for five nodes with an average fraction of native contacts
similar to the folding TS conformations but low connectivity/weight ratio
and/or high clustering coefficient.

\begin{figure}[h!]
\includegraphics[angle=-90,width=80mm]{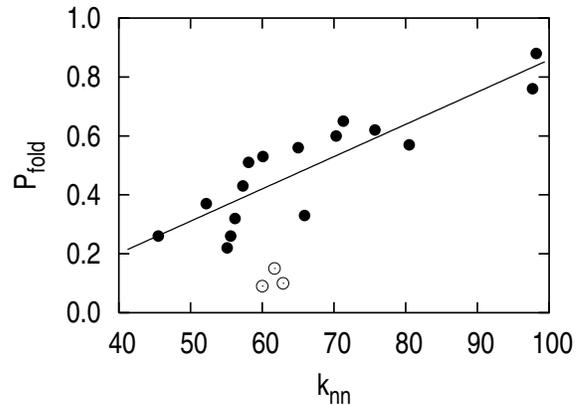}
\caption{Correlation between $P_{fold}$ and average neighbor
connectivity $k_{nn}$. Three nodes used as a control (low connectivity/weight
ratio and/or high clustering coefficient but similar fraction of native
contacts) are shown with empty circles.}
\end{figure}


\section{Conclusions}


Complex network theory was used to analyze the conformation space of a
structured peptide and the one of a random heteropolymer of same residue
composition. Four main results have emerged.  
First, as it was already observed for a variety of networks as diverse as the
World-Wide Web and the protein interactions in a cell, the conformation space
network of polypeptide chains is a scale-free network (power-law behavior of
the degree distribution).
Second, the native basin of the structured peptide shows a hierarchical
organization of conformations. This organization is not observed for the random
heteropolymer which lacks a native state.  
Third, free energy minima and their connectivity emerge from the network
analysis without requiring projections into arbitrarily chosen reaction
coordinates. As a consequence it is found that the denaturated state ensemble
is very heterogeneous and includes high entropy, high enthalpy conformations as
well as low entropy, low enthalpy traps.  
Fourth, the network properties were used to identify transition state
conformations and two main average folding pathways. It was found that the
average neighbor connectivity $k_{nn}$ correlates with $P_{fold}$, the
probability of folding.  $P_{fold}$ is computationally very expensive to
evaluate. Hence, it will be important to generalize this result by analyzing
other structured peptides which is work in progress in our research group.  In
conclusion, the network analysis seems particularly useful to study the
conformation space and folding of structured peptides including the otherwise
elusive transition state ensemble.

\vspace{0.5cm}\noindent{\bf Acknowledgments}\hspace{0.2cm} We thank 
M.\ Cecchini,
Prof.\ P.\ De Los Rios, 
E.\ Guarnera,
Dr.\ E.\ Paci,
Dr.\ M.\ Seeber and
Dr.\ G.\ Settanni
for interesting discussions.
This work was supported by the Swiss National Science Foundation
and the National Competence Center for Research (NCCR) in Structural Biology.

\bibliographystyle{elsart-num}
\bibliography{a-bib}


\clearpage


\setcounter{page}{1}
\renewcommand{\thepage}{S-\arabic{page}}

\setcounter{figure}{0}
\renewcommand{\thefigure}{S\arabic{figure}}

\begin{figure*}
{\Large\sc Supplementary Material}\\
\includegraphics[angle=0,width=170mm]  {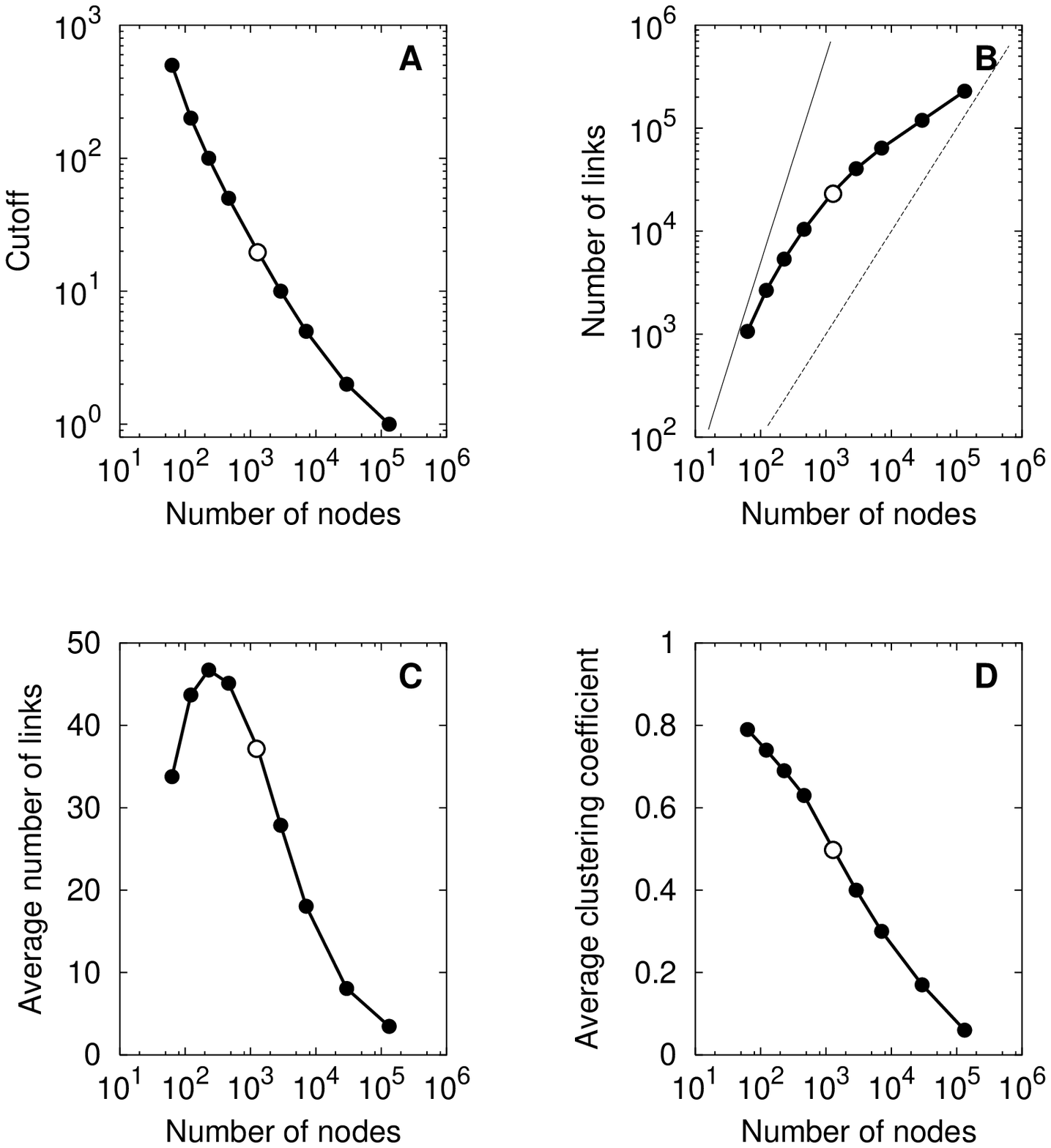} 
\caption{Dependence of the beta3s network properties on the node-weight
threshold. The threshold value used in the present work ($\tilde w=20$) is
shown as an empty circle while filled circles correspond to threshold values
of, from left to right, 500, 200, 100, 50, 10, 5, 2 and 1.  (A) Relation between
the threshold value and the number of nodes. (B) Number of links as a function
of the number of nodes.  When the threshold is very large (i.e., small number
of nodes) the network approaches a topology where all possible connections are
present (solid line, $N_{links}=N_{nodes}(N_{nodes}-1)/2$). When the threshold
is small (i.e., large number of nodes) the network approaches a topology with
only one link per node (dashed line, $N_{links}=N_{nodes}$). (C) Average number
of links per node $\left< k \right>$ as a function of the number of nodes. (D)
Average clustering coefficient $C$ as a function of the number of nodes.} 
\end{figure*}


\begin{figure*}
{\Large\sc Supplementary Material}\\
\includegraphics[angle=-90,width=170mm]  {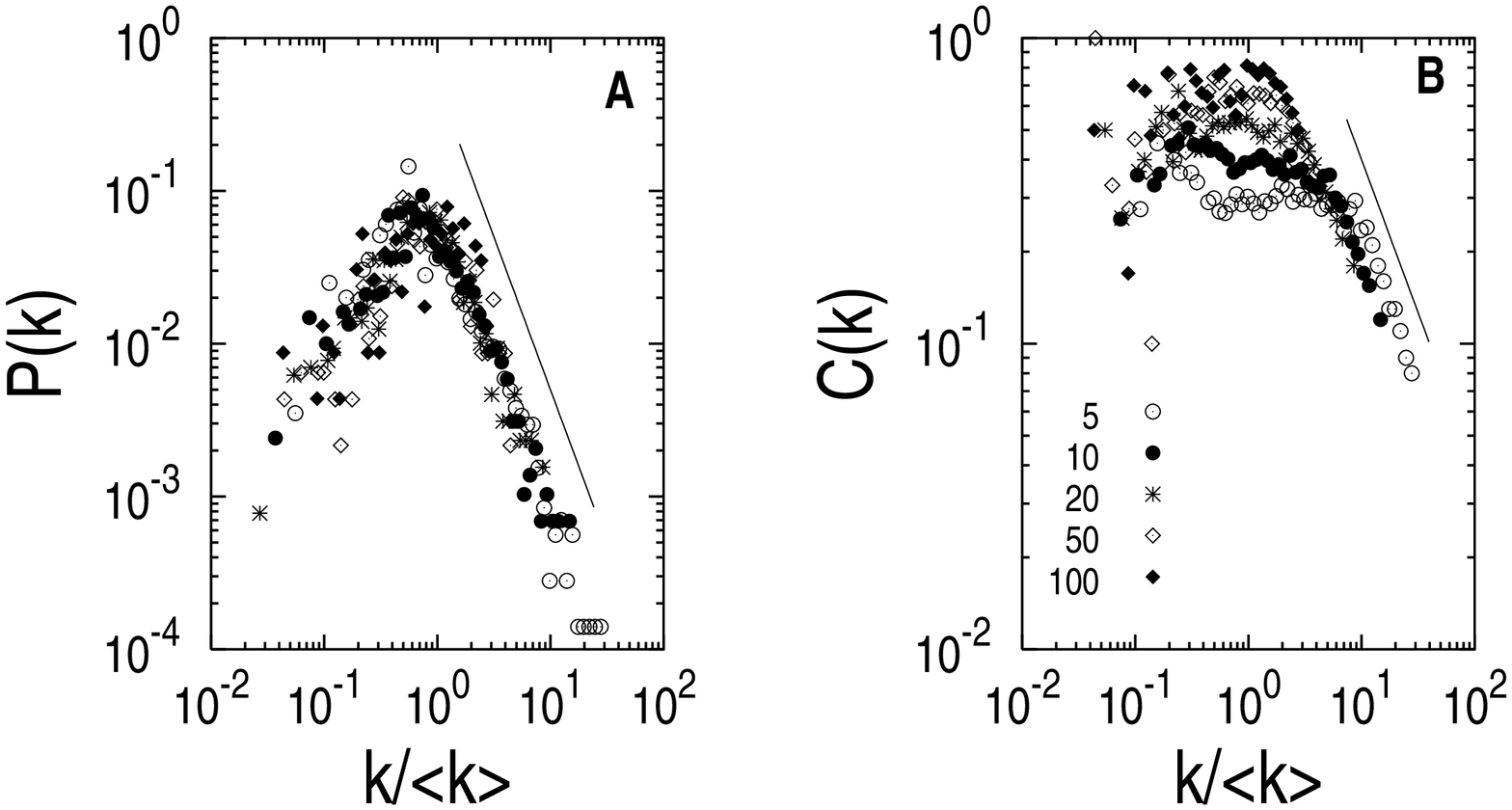}
\caption{Dependence of the beta3s connectivity distribution (A) and clustering
coefficient distribution (B) on the node-weight threshold.  This plot shows
that the scale-free behavior and the $1/k$ tail of the clustering coefficient
distribution are robust with respect to the choice of threshold values.}
\end{figure*}

\begin{figure*}
\includegraphics[angle=-90,width=170mm]  {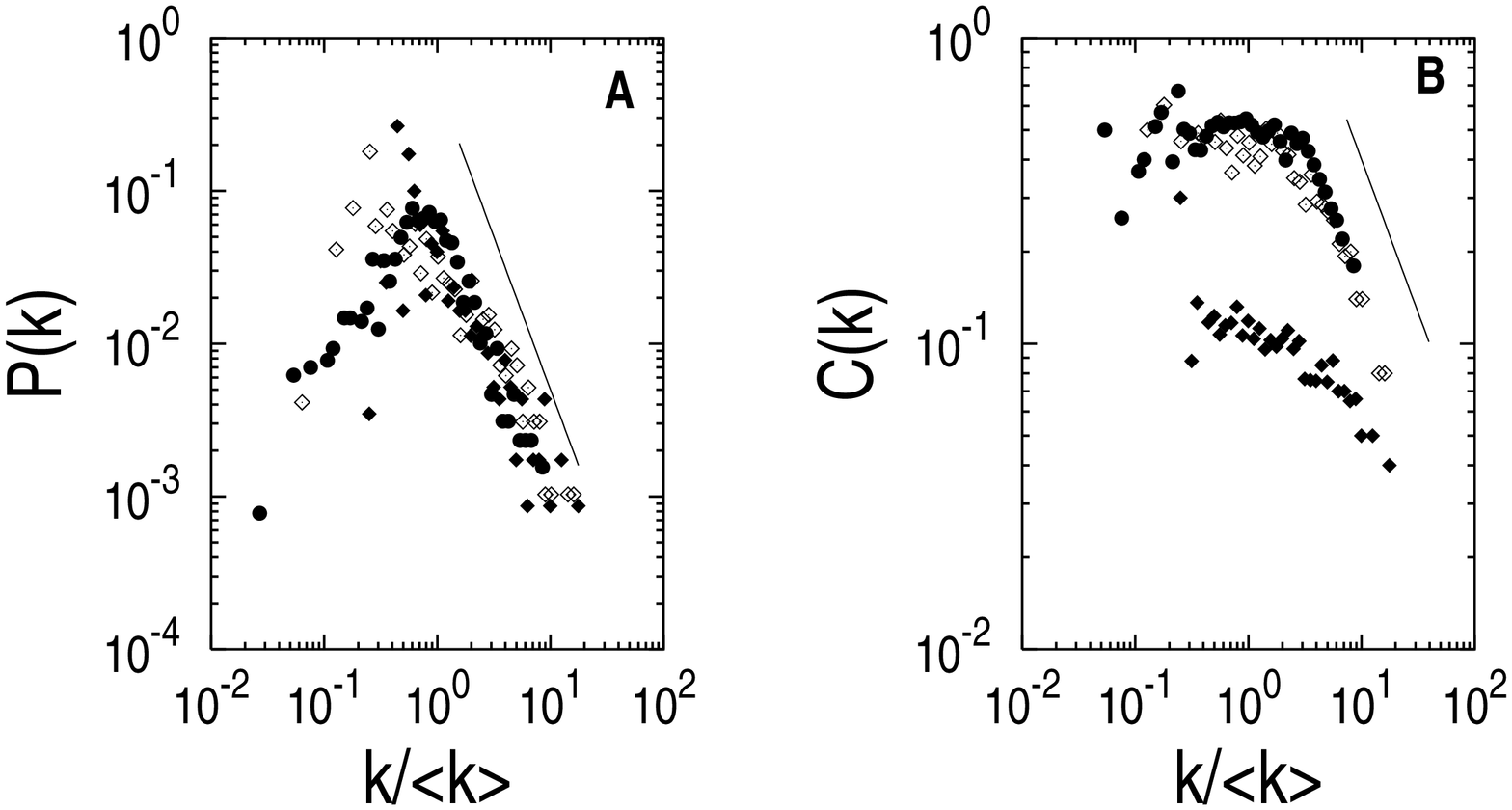}
\caption{Connectivity distribution (left) and clustering coefficient
distribution (right) for beta3s (filled circles), another structured peptide,
i.e., residues 101-111 of $\alpha$-lactalbumin (empty diamonds, Demarest et
al., (1999) {\it Biochemistry}, {\bf 38}, 7380), and a 20-residue homo-glycine
which is unstructured (filled diamonds).}
\end{figure*}

\setcounter{table}{0}
\renewcommand{\thetable}{S\arabic{table}}

\begingroup
\squeezetable
\begin{table*}[h]
{\Large\sc Supplementary Material}\\
\begin{ruledtabular}
  \caption{Supplementary material. Nodes used for $P_{fold}$ evaluation.}
  \begin{tabular}{ccccccccccc}
\parbox{1.0cm}{Node\\number} &
\parbox{1.0cm}{Probability\\of folding $P_{fold}$} &
\parbox{1.0cm}{Standard deviation\\$\sigma_{P_{fold}}$} &
\parbox{1.5cm}{Neighbor connectivity\\$k_{nn}$} &
\parbox{1.0cm}{Weight \\ $\tilde w$} &
\parbox{1.0cm}{Number of links\\$k$} &
\parbox{1.0cm}{$k/2\tilde w$} &
\parbox{1.0cm}{Clustering coefficient\\$C$} &
\parbox{1.0cm}{Native contacts\\$Q$} &
\parbox{1.0cm}{Standard deviation\\$\sigma_Q$} &
\parbox{1.5cm}{Secondary\\ structure string} \\ \\
  \hline
432 & 0.22 & 0.17 & 55.1 & 54  & 40  & 0.37 & 0.31 & 	0.38 & 0.09 & {\tt -----SS-EEEEESSEEEE- } \\
218 & 0.26 & 0.20 & 45.5 & 105 & 73  & 0.35 & 0.23 &	0.42 & 0.09 & {\tt ----SSS---EEESSEEE-- } \\
313 & 0.26 & 0.16 & 55.6 & 70  & 60  & 0.43 & 0.28 & 	0.46 & 0.09 & {\tt -EEE-STTEEE-SSS----- } \\
446 & 0.32 & 0.25 & 56.2 & 52  & 50  & 0.48 & 0.28 & 	0.40 & 0.10 & {\tt --EEESSEEE---SS----- } \\
308 & 0.33 & 0.28 & 65.9 & 72  & 69  & 0.48 & 0.23 & 	0.44 & 0.08 & {\tt ----SSS--EEEESSEEEE- } \\
315 & 0.37 & 0.26 & 52.2 & 70  & 60  & 0.43 & 0.24 & 	0.47 & 0.08 & {\tt -EEE-STTEEE--SS----- } \\
306 & 0.43 & 0.27 & 57.3 & 73  & 60  & 0.41 & 0.31 & 	0.42 & 0.08 & {\tt -----SS--EEE-STTEEE- } \\
208 & 0.51 & 0.26 & 58.1 & 115 & 87  & 0.38 & 0.23 &	0.43 & 0.08 & {\tt -----SS--EEEESSEEEE- } \\
589 & 0.53 & 0.31 & 60.1 & 40  & 52  & 0.65 & 0.17 & 	0.45 & 0.10 & {\tt ---SSSTT-EEEESSEEEE- } \\
580 & 0.56 & 0.34 & 65.0 & 40  & 47  & 0.59 & 0.26 & 	0.48 & 0.08 & {\tt -EEEESSEEEE--SSS---- } \\
197 & 0.57 & 0.39 & 80.5 & 121 & 105 & 0.43 & 0.28 &	0.52 & 0.09 & {\tt -----STT-EEEESSEEEE- } \\
540 & 0.60 & 0.16 & 70.3 & 44  & 49  & 0.56 & 0.28 & 	0.46 & 0.07 & {\tt -----GGG-EEE-STTEEE- } \\
285 & 0.62 & 0.30 & 75.7 & 80  & 68  & 0.42 & 0.30 & 	0.47 & 0.07 & {\tt -----GGG-EEEESSEEEE- } \\
630 & 0.65 & 0.22 & 71.3 & 38  & 56  & 0.74 & 0.29 & 	0.44 & 0.11 & {\tt -----STT--EE-STTEE-- } \\
426 & 0.76 & 0.20 & 97.7 & 55  & 76  & 0.69 & 0.43 & 	0.55 & 0.12 & {\tt ---B-TTTB-EEESSEEE-- } \\
280 & 0.88 & 0.18 & 98.2 & 82  & 81  & 0.49 & 0.43 & 	0.56 & 0.09 & {\tt ---EESSEE-EE-STTEE-- } \\
Control simulations \\
174 & 0.09 & 0.10 & 60.0 & 139 & 51  & 0.18 & 0.61 &	0.37 & 0.07 & {\tt --EE-STTEEEESTTEEEE- } \\
179 & 0.09 & 0.10 & 25.0 & 135 & 19  & 0.07 & 0.33 &	0.44 & 0.07 & {\tt --EESSEE--EEESSEEE-- } \\
15 & 0.10 & 0.13 & 35.8 & 3243 & 73  & 0.01 & 0.28 &	0.47 & 0.08 & {\tt --EEESSSEEEEESSEEEE- } \\
200 & 0.10 & 0.27 & 62.9 & 119 & 51  & 0.21 & 0.31 &	0.31 & 0.07 & {\tt ----BSSB--EEESSEEEE- } \\
475 & 0.15 & 0.17 & 61.7 & 48  & 34  & 0.35 & 0.68 & 	0.43 & 0.06 & {\tt -EEE-STTEEETTTT-EEE- } \\
\end{tabular}
\end{ruledtabular}
\end{table*}
\endgroup

\end{document}